\documentclass{article}
\usepackage{waspaa21,amsmath,graphicx,url,times}
\usepackage{color}

\title{Identifying Actions for Sound Event Classification}

\name{Benjamin Elizalde$^{\dagger}$, Radu Revutchi, $^{\star}$Samarjit Das, Bhiksha Raj, Ian Lane, Laurie M. Heller 
\vspace*{-5pt} 
\thanks{$^{\dagger}$Submitted while at Microsoft, benjaminm@microsoft.com. Thanks to Bosch Research, CONACyT and the SOWG for their financial support.}} 
\address{Carnegie Mellon University, $^{\star}$Bosch Research Pittsburgh \\
Email: {bmartin1,radurevutchi,bhiksha,ianlane,laurieheller}@cmu.edu, samarjit.das@us.bosch.com
\vspace*{-10pt}
}

\begin{document}

\ninept
\maketitle

\begin{sloppy}

\begin{abstract}
\vspace*{-0.07in}
In Psychology, actions are paramount for humans to identify sound events. In Machine Learning (ML), action recognition achieves high accuracy; however, it has not been asked whether  identifying actions can benefit Sound Event Classification (SEC), as opposed to mapping the audio directly to a sound event. Therefore, we propose a new Psychology-inspired approach for SEC that includes identification of actions via human listeners. To achieve this goal, we used crowdsourcing to have listeners identify 20 actions that in isolation or in combination may have produced any of the 50 sound events in the well-studied dataset ESC-50. The resulting annotations for each audio recording relate actions to a database of sound events for the first time. The annotations were used to create semantic representations called Action Vectors (AVs). We evaluated SEC by comparing the AVs with two types of audio features -- log-mel spectrograms and state-of-the-art audio embeddings. Because audio features and AVs capture different abstractions of the acoustic content, we combined them and achieved one of the highest reported accuracies (88\%).
\end{abstract}

\vspace*{-0.1in}
\section{Introduction}
\label{sec:intro}
\vspace*{-0.1in}
Sound Event Classification (or Audio Tagging) aims to assign a sound event class label to an audio clip (e.g., dog, car horn). Typically, SEC consists of taking an input audio file, computing features inspired by auditory perception, and using them to train ML classifiers. Motivated by the success of understanding how humans hear in order to build acoustic intelligence, we looked further into human perception and identification of sound events. 

Psychomechanics is the study of the perception of physical properties of sound-sources~\cite{lemaitre2018acoustics}, which has found that actions distinguish sound events better than the source materials, size and shape. Gaver~\cite{gaver1993world} argued that listening to everyday sounds focuses mainly on the physical aspects producing the sound. Lemaitre and Heller~\cite{lemaitre2012auditory} concluded that the listener's ability to identify solid materials based on audio is in general not accurate, except for resonant vs nonresonant impacts  (e.g. glass vs. plastic). Moreover, perception of size and shape are in general less reliable than material~\cite{grassi2005we, grassi2013looking, houben2005contribution, lutfi2001auditory}. In contrast, Lemaitre and Heller~\cite{lemaitre2013evidence} showed that actions are a robust property perceived in everyday sounds. They also found that actions producing simple sound events were better identified than the materials~\cite{lemaitre2012auditory}. Verbs can describe actions and interactions between objects and sometimes also the material the objects are made of~\cite{darvishi1995designing, Schafer1993, Gygi2004, ballas1987interpreting, dubois2006cognitive}. VanDerveer~\cite{vanderveer1980ecological} found that people who were asked to identify sounds made by common objects would spontaneously describe the actions involved in generating the sounds. Hence, it would be interesting to see if the importance of actions to humans provides a cue for improving ML models.

In fact, despite the scarce literature of ML models trained on physical properties of sound-sources, models trained on audio labeled with actions and materials have achieved successful accuracy. Owen et al~\cite{owens2016visually} recorded videos of hitting a stick on surfaces of different materials (e.g., grass, metal, water, wood) and achieved audio-based classification of eleven materials with 45.8\% accuracy. Previously in ~\cite{sager2018audiopairbank}, we collected audio labeled with 400 suffixed nouns derived from verbs referring to the action generation (e.g., ``clapping crowd"). Then, we evaluated binary classification for each suffixed noun and achieved an overall 71\% accuracy (chance was 50\%), with classes performing as high as ``ringing alert” with 92\%. In Psychology, identifying actions has been a successful intermediate step for recognizing sound events, yet in ML there is not enough evidence that a similar approach would benefit SEC.

Identifying actions for SEC can potentially be used across different sound event datasets and categories. Typically, SEC processes the audio signal and maps it to a sound event. Actions can serve to bridge the audio with the sound event, providing an interpretable step to explain SEC. This is important because some sound event datasets have many categories named after sound-sources despite the fact that the same object can produce different types of sounds. Furthermore, the dataset of ESC-50~\cite{ESC50} demonstrated that for listeners and ML models, classification of ``airplane" and ``helicopter" were highly confused because they both share similar sounds produced by the propeller and rotor engine respectively. In this case, identifying the sound produced by the action of rotating could serve to explain the acoustic confusion, because actions tend to produce consistent acoustics that can cut across the semantics of sound events labels. It would also help to determine if we can combine audio examples of the same class from different datasets or annotation processes for the purposes of training~\cite{elizalde2013audio}. This would require a set of actions that could describe a set of sound events with a larger number of classes in different contexts.

Therefore, we propose a new Psychology-inspired approach for SEC that includes identification of actions. We used crowdsourcing to identify 20 actions that in isolation or in combination produced any of the 50 sound events in the well-studied dataset ESC-50. The resulting annotations for each audio recording relate actions to sound events for the first time (available online). Combining audio features and action identification improved our SEC. This demonstrates a benefit of drawing from domain knowledge within Psychology to design ML algorithms for SEC. 
\vspace*{-0.1in}
\section{Relating Actions to Sound Events}
\label{sec:relating}
\vspace*{-0.1in}

In order to relate actions to sound events, we chose a well-studied sound event dataset named ESC-50~\cite{ESC50}. ESC-50 has 50 classes from five broad categories: animals, natural soundscapes and water sounds, human non-speech sounds, interior/domestic sounds, and exterior sounds. Each class has 40 five-second-long sounds for a total of 2,000 audio files. The categories do not necessarily have intra-class acoustic consistency. The sound events are generally exposed in the foreground with limited background noise. Field recordings may exhibit overlap of competing sound sources in the background. The audio comes from the Internet, hence the recording process is unknown. Category-labels consist of one or two words; only about 14\% are labeled with a specific action and most are labeled with nouns. The next step was to select relevant actions.

Selecting actions that in isolation or combination produced all of the ESC-50 sound events was challenging. The recording process and definition of the sound event classes were unknown; thus, we listened to audio recordings corresponding to each class and chose actions that could have produced at least part of the audio content. In order to distinguish our approach from the existing labels, we chose actions that did not result in simply relabelling a sound event class with an action-oriented synonym for its existing object-focused label. For example, choosing the action ``sawing" would have been a one-to-one relabelling of the sound event class ``hand sawing". This approach allowed us to draw new insights from action classes that might supplement existing ML systems. The number of actions determined the number of options that would be given to the annotators; this number was kept moderate in order to keep the task feasible and avoid fatigue for participants. Our 20 actions were in line with prior listening studies for action identification ~\cite{lemaitre2013evidence} which also utilized about 20 category-options at a time. 

The selection of actions was inspired by listening experiments in the literature of Psychology and our own listening experiments in our auditory lab. We drew actions from a previously collected dataset by Heller~\cite{hellerabstrract} and two taxonomies of everyday sounds, Gaver’s~\cite{gaver1993world} and Houix’s~\cite{houix2012lexical}. The taxonomies included different varieties of actions, such as interactions (friction), specific actions (scraping), manners of actions (scraping rapidly), and objects of actions (scraping a board). While  actions  can be described in general terms (such as impact), Lemaitre and Heller~\cite{lemaitre2013evidence} showed that the most specific descriptors (such as tap) produce the highest accuracy and fastest identification response (analogous to superordinate vs. subordinate categories of objects such as furniture vs. chair). Thus, we constrained our study to 20 specific actions that were not redundant, were not strictly associated with one class, and ensured that each class was associated with at least one action. The actions Table~\ref{table:actions} were used to annotate all 2000 audio files in ESC-50.

We did not distribute the action verbs to match exact proportions of sound classes in ESC-50 in any possible way that they could be divided up (e.g. by source material, source shape, living/nonliving, etc.), but their distribution is not far from proportionate. For example, we have 25\% of actions related to voice for about 32\% of vocalization-related sound events. We have 15\% of actions related to liquids (dripping, splashing, pouring) for about 16\% of liquid-related sound events.

\begin{table}[ht]
	\centering
	\resizebox{3in}{!}{%
	\begin{tabular}{l|l|l|l|l}
		dripping & splashing & pouring & breaking & calling \\  
		rolling & scraping & exhaling & vibrating & ringing \\  
		groaning & gasping & singing & tapping & wailing   \\ 
		crumpling & blowing & exploding & rotating & sizzling \\
	\end{tabular}%
	}
	\caption{We selected 20 actions that in isolation or combination could have produced at least part of the 50 sound events.}
	\label{table:actions}
\end{table}

Due to the large number of audio files (2,000) in ESC-50 dataset, we identified actions using crowdsourcing (Mechanical Turk). We designed an interface that included a playable audio clip (without showing the sound event label), the prompt ``For each action below, judge how likely it is to have produced at least part of the sound event.", followed by the 20 actions to be scored. The scores were inspired by a five-point Likert scale ranging from 0-4 where 0 meant that the action contribution was very unlikely and 4 that it was very likely. We asked for annotators to be fluent in English, wear headphones, have no hearing impairments, and be between 18 and 60 years old. We selected participants using the filters: Task (HIIT) Approval Rate greater than 95\% and a number of HIIT approved greater than 1000. Our rejection criteria discarded all annotations from any individual who gave a 3 or 4 to the vast majority of actions for the same sound. We had high-quality in-lab annotated recordings on a subset of ESC-50 that helped to validate our rejection criteria. We measured the inter-annotator agreement between the lab and MTurk annotations using Fleiss Kappa score resulting in Fair Agreement. Each audio file was annotated with each of the 20 actions by three different participants for a total of 6,000 x 20 ratings in the entire dataset\footnote{{GitHub: bmartin1/Identifying\_Actions\_for\_Sound\_Event\_Classification}}. 

\vspace*{-0.1in}
\section{Experiments and Results}
\label{sec:experiments}
\vspace*{-0.1in}

\subsection{Creating Action Vectors}
\label{sec:avs}
\vspace*{-0.08in}
After collecting the annotations derived from identifying actions in ESC-50, we created Action Vectors (AVs). An AV is a 20-dimensional representation where each dimension (dim) is an action and the score in each dimension determines how likely the action was to have contributed to the sound event. One AV from one rater has 20 dims each of which ranges from 0-4. To create an AV for each audio file, we summed the scores across the three annotators. Although we didn't explicitly express time in the AVs in the form of temporal frames, some temporal properties are implied by actions; for example, intermittent vs continuous audio is implied by tapping vs rolling actions. 
\begin{figure}[h]
   \centering
     \includegraphics[width=0.48\textwidth]{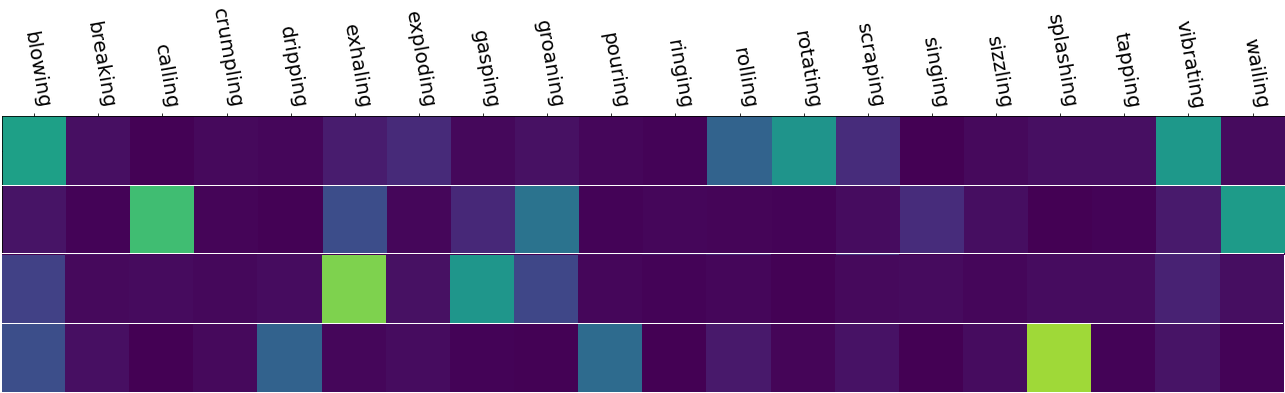}
     \caption{Rows: average AVs for airplane, cat, coughing, and sea waves. Columns: action ratings. Brighter green = more likely.}
     \label{fig:av_heatmap}
\end{figure}
Although the AVs varied among exemplars within a class, the average pattern for each class yields a unique pattern of ratings (in a 50x20 matrix). For example, the highest action ratings shown in Fig.~\ref{fig:av_heatmap} differed between airplanes (blowing, rotating, and vibrating), cats (calling, groaning, and wailing), coughing (exhaling and gasping), and sea waves (dripping, pouring, and splashing)\footnote{Listed if at least 1 standard deviation above the mean.} Thus, the multidimensional AV for each file does not equate to a single verb or class. Once we created 2,000 AVs for ESC-50, we used them for SEC. 
\vspace*{-0.08in}
\subsection{Comparing Sound Event Classification using Audio Features and Action Vectors}
\label{sec:comparing}
\vspace*{-0.08in}

We study for the first time how identifying actions can benefit SEC, as opposed to mapping the audio directly to a sound event. Whether AVs could improve performance with a straightforward setup was not obvious. Hence, we compared SEC utilizing AVs vs. two common audio features. The two SEC pipelines for our experiments are in Fig.~\ref{fig:av_pipeline}. On the left is the typical approach that takes the input audio, computes features and assigns a class label. On the right, we depict our addition of an intermediate step where listeners identify actions in the audio. 

\begin{figure}[t]
   \centering
     \includegraphics[width=0.38\textwidth]{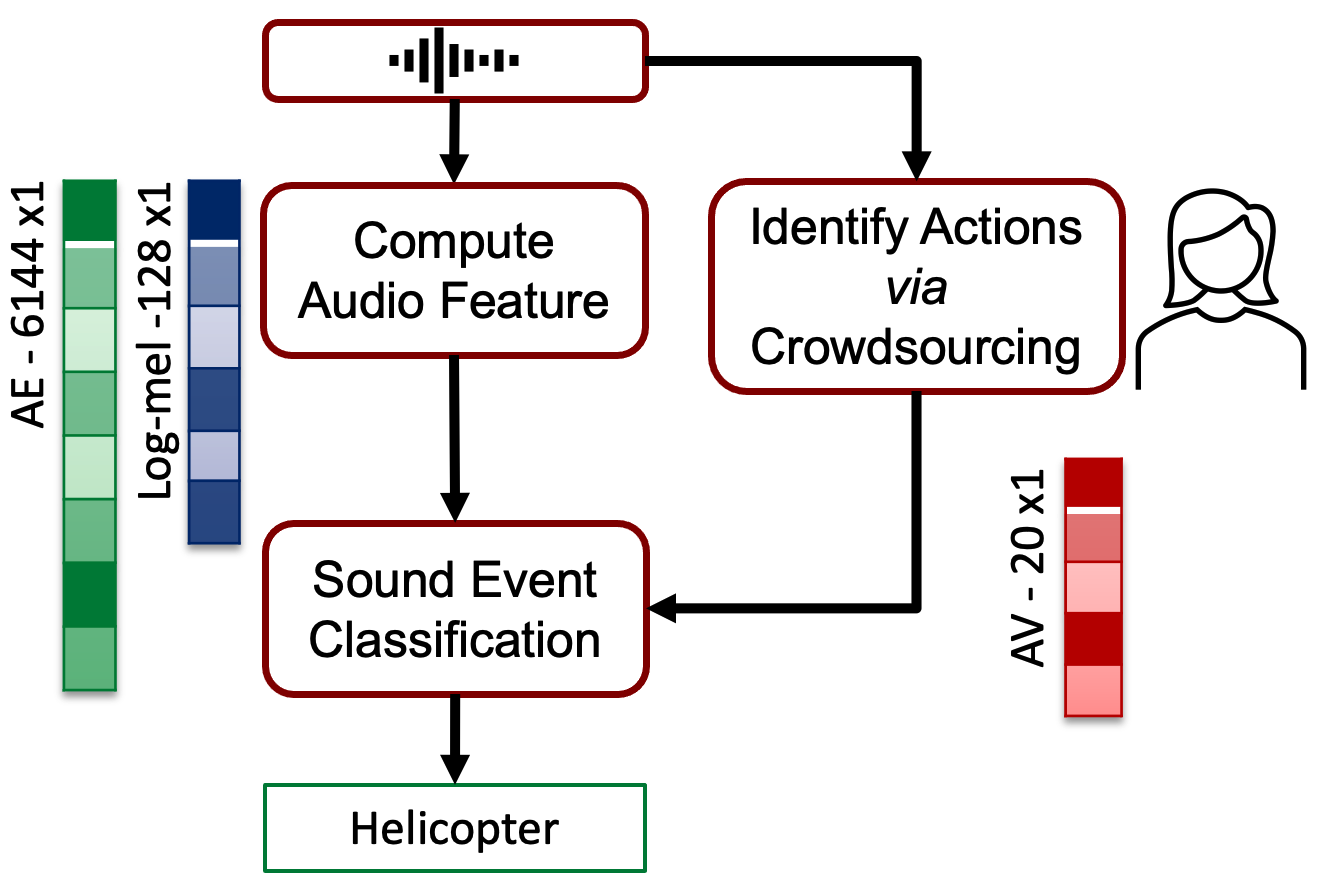}
     \caption{Typically, SEC processes the input audio and assigns a class. We propose to add an intermediate step where listeners identify actions in the audio. The actions are transformed into Action Vectors and are used for automatic SEC.}
     \label{fig:av_pipeline}
\end{figure}

For the input audio, we computed two commonly-used types of audio features, log-mel spectrograms and state-of-the-art data-driven audio embeddings (AEs). We computed spectrograms using the librosa package~\cite{mcfee2015librosa} with default settings resulting in $128$ mel-filters summarized with their mean across time frames. We computed AEs using a network called OpenL3~\cite{Cramer:LearnMore:ICASSP:19}, which was pre-trained with $60M$ videos. The parameters used were: content\_type: $music$, input\_repr: $mel256$, embedding\_size: $6144$. The AEs were summarized with their mean across time frames. The features and the AVs were normalized ($L2$) by removing the means and scaling to unit variance before being passed to the classifier (AEs+AVs did not include the second step). We summarized our features over time because it is still common in the literature~\cite{Cramer:LearnMore:ICASSP:19} and it is not trivial to combine unaveraged representations with unidimensional AVs. 

The features and the AVs were used to train two types of ML classifiers for SEC, a linear Support Vector Machine (SVM) and a Deep Neural Network (DNN). The SVM classifier provides fast computation and a basic model. We included a DNN as a non-linear classifier alternative that often performs better than linear classifiers. The hyper-parameters of the classifiers were tuned. For the SVM we set the soft-margin to $C=35$, one-vs-rest multiclass algorithm optimized with the primal approach (scikit-learn~\cite{scikit-learn}). The DNN had 5 linear layers: the first is from number of features to 800, then from 800 to 500, then from 500 to 200, and then from 200 to number of classes. Each hidden layer had $tanh$ activation. The output layer had a $softmax$ activation combined with a Categorical Cross Entropy Loss. We used SGD optimizer ($lr=0.008$) and $100$ epochs. As ESC-50 distributes the audio files into 5 folds, we ran the pipeline using all combinations in which each fold was the test set once. Overall accuracy was computed across the five folds to evaluate SEC performance. The SVM produces the same accuracy in every run, but the DNN was run 10 times to compute the mean and standard deviation ($\sigma$).

SEC using AVs resulted in good performance, suggesting that AVs carry useful information to bridge audio and sound events. Table~\ref{table:SEC_AVs} shows SEC accuracy using AVs and both types of audio features. SEC with AVs yielded 48.25\% with the SVM and 51.81\% ($\sigma$=0.4\%) with the DNN. SEC with log-mel spectrograms yielded 30.70\% with the SVM and 34.00\% ($\sigma$=0.3\%) using the DNN, which is consistent with other papers that use log-mel-based features~\cite{ESC50}. SEC with AEs yielded 80.9\% with the SVM and 81.46\% ($\sigma$=0.3\%) with the DNN, which is consistent with the OpenL3 paper~\cite{Cramer:LearnMore:ICASSP:19}. The AVs outperformed log-mel spectrograms by an absolute 15\% to 17\% accuracy, but AVs under performed AEs by an absolute 30 to 28\% accuracy. 
\begin{table}[t]
	\centering
	\resizebox{2.7in}{!}{%
	\begin{tabular}{l|c|c}
		\textbf{Input Features}   & \textbf{linear SVM} & \textbf{DNN}  \\  \hline  
		 log-mel spectrograms              & 30.70\%  & 34.00\% \\ 
		 AVs (Action Vectors)              & 48.25\%  & 51.81\% \\ 
		 AEs (Audio Embeddings)            & 80.90\%  & 81.46\% \\           
		 AEs + log-mel                     & 74.35\%  & 78.77\% \\           
		 AVs + log-mel                     & 55.05\%  & 69.50\% \\  
		 \textbf{AVs + AEs }               & \textbf{86.60\%}  & \textbf{88.00\%} \\           
		 AVs + AEs + log-mel               & 78.35\%  & 83.31\% \\           
	\end{tabular}%
    }
	\caption{Classification accuracy with different inputs. AEs+AVs achieved one of the highest accuracy reported in ESC-50.}
	\label{table:SEC_AVs}
	\vspace*{-0.2in}
\end{table}

We combined AVs and audio features independently to evaluate how they complement one another for SEC. We used an SEC pipeline similar to the one used in the first set of experiments and tried four combinations: AVs with log-mel spectrograms, AVs with AEs, log-mel spectrograms with AEs, and all together. 

SEC combining identification of actions and audio features resulted in one of the highest accuracies reported in ESC-50. Table~\ref{table:SEC_AVs} shows SEC using the four concatenated combinations. SEC with AVs+AEs yielded 86.6\% accuracy with SVM and 88.00\% ($\sigma$=0.4\%) with DNN. In the literature, only a few approaches surpass human performance of 81.3\%~\cite{ESC50}.

Our SEC approach with AVs+log-mel spectrograms yielded 55.05\% with SVM and 69.50\% ($\sigma$=0.3\%) with DNN. In both cases the combination resulted in better performance than in isolation. On the other hand, stacking features does not always translate to better performance, as exemplified by log-mel spectrograms in our study (although they may still show benefits in other implementations): SEC with stacked AEs+log-mel spectrograms yielded 74.35\% with SVM and 78.77\% ($\sigma$=0.5\%) with DNN and AVs+AEs+log-mel yielded 78.35\% accuracy with SVM and 83.31\% ($\sigma$=0.2\%).


While it is not surprising if adding more feature types and dimensions improves performance, it is remarkable that by adding 0.3\% dims (20/6144) to the AEs we can improve SEC performance by almost an absolute 7\%. It is also interesting to see that with a vocabulary of 20 actions we can express the diversity of a set of 50 sounds from different types (e.g. human, animals, environments, machines, etc). The finding that the information was complementary was not known at the start of this project and we feel that this is a primary contribution. While it seemed logical that AVs would capture information relevant to identify sound events, we didn't know if it was going to be redundant with an already strong audio representation such as the AEs.

\vspace*{-0.1in}
\section{Discussion}
\label{sec:discussion}
\vspace*{-0.1in}

The use of actions represented by AVs has benefits, such as: providing novel information via the graded combination of multiple actions per sound, providing a basis for understanding the confusability and heterogeneity of certain sound classes, providing an interpretable semantic representation, and providing a new way to relate sounds. We investigated whether the main reason why AVs provide a benefit is because multiple actions are combined to characterize a sound event instead of labelling sound events with one dominant action. To test this idea, we scaled the scores of the AVs to be from 0 to 1 (instead of 0 to 12) and then quantized them with a threshold of $0.5$. Scores under the threshold were set to $0$, and greater or equal than the threshold were set to $1$. Then, we ran SEC concatenating AEs and the new quantized AVs and achieved 82\% instead of 86.75\% (DNN). This means that there is information in the graded ratings of multiple actions per sound. On average, AVs have non-zero scores in $6$ out of $20$ dims. Hence, our approach offers more nuanced information than would a simple replacement of object-based class labels with equivalent action-based verb labels.

We have not attempted to equate dimensions across different features because our goal is not to put these methods head to head, but rather to see if collectively they can improve performance when put together. Although we didn't do a formal analysis of the dimensionality, if anything we are underestimating the power of our AVs because we are comparing them against features with one and two orders of magnitude greater dimensionality and more information. We were interested in the low dimensional AVs to see if they have any representational power, and we showed that they do. Nonetheless, here we compare dims for reference. AVs with only $20$ dims can be helpful to classify 50 sounds, whereas the log-mel spectrograms have $128$ dims and the AEs have $6144$. As an illustration of the effect of reducing dims for a given feature, when extracting the AEs with only $512$ dims (a parameter option in OpenL3) SEC accuracy dropped from 79.75\% to about 40\%. Holding the hyperparameters constant in the DNN, the AEs with $512$ dims underperformed the AVs with $20$ dims (51.81\%). Depending on the application, it is potentially useful that each dim in the AVs is semantically interpretable as opposed to the latent meaning in AEs.

Whether AVs are part of an SEC pipeline or not, they can help to explain inter- and intra-class confusion. For instance, the high confusion occurring between ``airplane" and ``helicopter" can be explained by a subset of ``airplane" recordings having sounds produced by a propeller-powered engine, which are similar to the sounds produced by the rotor engine of a helicopter. These recordings scored high for the action ``rotating". Actions tend to produce consistent acoustics that can cut across the semantics of sound events. A set of actions can describe sound events with similar acoustic content, but with different labels and in different contexts. We could also use ``rotating" for the sound event of ``washing machine" in domestic sounds. 

We expect that our selection of actions should work with other relevant sound event datasets. The selected actions are inspired by ESC-50, which has a large diversity of sounds. We stayed away from using actions that were essentially identifications of the sound sources (e.g. ``mooing" for ``cow") and aimed to use verbs that could apply more broadly. We did not include verbs that did not apply to ESC-50 so as to reduce the amount of hand-labelling required. Hence, we did not have any words specifically designed to lump together objects that produced intra-class variability, such as ``flying" or ``airplane", so the intra-class variability wouldn't be unique to ESC-50. For example, the same variability pointed out in the previous paragraph happens in Google's AudioSet (over 500 classes and over a million recordings), where from 66 ``fixed-wing aircraft"~\footnote{\url{research.google.com/audioset//eval/fixedwing_aircraft_airplane.html}} recordings, some are produced by the jet engines and some by propeller-powered engines. We expect that our selection of actions could work with a subset of AudioSet, but a complete study was out of the scope of this paper. To increase the SEC accuracy and interpretable capabilities of AVs we should increase the number of actions.

SEC accuracy using AV depends on the selection and number of actions. Not all of the actions affect SEC equally. This is determined by how a combination of actions can distinguish sound events. For example, one of the actions was ``calling", which scored high in $11$ sounds produced by vocal tracts, such as ``frog", ``dog", ``crow", resulting in inter-class confusion. When removing those $11$ classes from the dataset, SEC accuracy with AEs remained around 80\%, but with AVs increased from 51\% to 57\%. By adding more actions that distinguish vocalizations, the better AVs can discriminate between such sounds events. The selection of actions also affects how we organize sound events.

\begin{figure}[ht]
   \centering
     \includegraphics[width=0.35\textwidth]{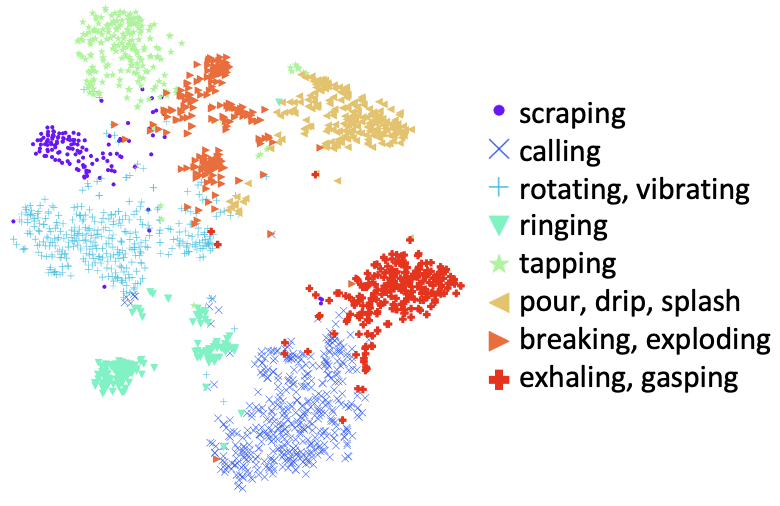}
     \caption{AVs provide a new way to relate ESC-50 (or any other) sound events based on shared actions.}
     \label{fig:av_clusters}
\end{figure}

AVs provide a new way to relate sound events based on shared actions. We grouped the $2,000$ AVs of ESC-50 using K-means (k=8) and plot them (tsne, perplexity=50) in Fig.~\ref{fig:av_clusters}. Each of the $8$ groups was assigned a label corresponding to its dominant action(s): scraping, calling, rotating and vibrating, ringing, tapping, pouring and dripping and splashing, breaking and exploding, exhaling and gasping. Varying the number of clusters created groups with different dominant actions. Actions and physical properties can be used to build knowledge of sound events~\cite{guastavino2018everyday,elizalde2020never} and automatic description of both could help label audio samples without having to force them into a single class, which is particularly helpful when processing unlabeled audio in the large-scale~\cite{elizalde2020never}.

To improve SEC accuracy, usability of actions for different datasets, and description of sounds, scaling the number of actions is a logical direction, but it poses some challenges. Our action annotations could be used to train ML models that take an audio recording and generate the AV automatically. The challenging part is to design an ML model that can generate AVs with graded ratings similar to the hand-labeled AVs. Another potential scaling approach is to train independent action recognition models and then construct the AVs. We hope our findings in this paper can inspire the community to explore action recognition for SEC and description of sound events via physical properties.
\vspace*{-0.1in}
\section{Conclusions}
\label{sec:conclusions}
\vspace*{-0.1in}
We demonstrated that our proposed Psychology-inspired approach of identifying actions improved SEC. AVs were derived from humans identifying actions, but we are in the process of using our annotations to train models that can automatically generate the AVs given an audio file. We found a benefit from using human expertise to describe sound events with a suite of actions rather than forcing each sound to be represented by only one action.

\bibliographystyle{IEEEtran}
\bibliography{2021_WASPAA_PaperTemplate_Latex}
%
%
%
%
%
%
%
%
%

\end{sloppy}
\end{document}